\documentclass[11pt,twocolumn]{article}
\usepackage{amsmath}
\usepackage{amssymb}
\usepackage{amsthm}
\usepackage{epsfig}
\usepackage{hyperref}
\usepackage[margin=0.7in]{geometry}
\usepackage{cite}
\usepackage{graphicx}
\usepackage{epstopdf}
\usepackage{color}

\title{Self-organization in a distributed coordination game through heuristic rules}
\author{
Shubham~Agarwal\footnote{Email: shubham119413@gmail.com, Indian Institute of Technology, Chennai 600036, India.}
\and
Diptesh~Ghosh\footnote{Email: diptesh@iima.ac.in,
Production \& Quantitative Methods Area, Indian Institute of Management, Ahmedabad 380015, India.}
\and Anindya~S.~Chakrabarti\footnote{(corresponding author) Email : anindyac@iima.ac.in,
Economics Area, Indian Institute of Management, Ahmedabad 380015, India.}
}

\begin{document}

\maketitle

\begin{abstract}
In this paper we consider a distributed coordination game played by a large number of agents with finite information sets, which characterizes emergence of a single dominant attribute out of a large number of competitors.  Formally, $N$ agents play a coordination game repeatedly which has exactly $N$ Nash equilibria and all of the equlibria are equally preferred by the agents. The problem is to select one equilibrium out of $N$ possible equilibria in the least number of attempts. We propose a number of heuristic rules based on reinforcement learning to solve the coordination problem. We see that the agents self-organize into clusters with varying intensities depending on the heuristic rule applied although all clusters but one are transitory in most cases. Finally, we characterize a trade-off in terms of the time requirement to achieve a degree of stability in strategies and the efficiency of such a solution.
\end{abstract}


\vskip .5cm


\section{Introduction}

Understanding collective behavior of large-scale multi-agent systems is an important question in the econophysics and the sociophysics literature \cite{Namatame_16,Sen_Chakrabarti_14}. Often in social and economic worlds, we find emergence and evolution of global characteristics that cannot be explained in terms of fundamental properties \cite{Watts_culture_06}.  We find examples of particular social norms or technologies that become more popular than their competitors, which are not necessarily worse in terms of attributes. 
Similarly, norms and opinions emerge as an equilibrium through reinforcement among the social and economic agents \cite{Watts_opinion_07}. Leaders emerge in the political context through a complicated process of competition and interaction among millions of individuals \cite{Stauffer_06}. In this paper, we present a simple multi-agent game to study the emergence of
one dominant attribute out of many potential competitors through complex and adaptive interactive processes (\cite{Namatame_16};see also Ref. \cite{PuglieseEJB09}).

We focus on two properties of large scale interaction. One, agents can coordinate to specific choices from a number of potentially identical choices which may also be interpreted as emergence of cooperation \cite{Challet1997}
and two, such coordination may take time to arrive at but once arrived, can be quite stable. 
Therefore, we address the dynamic (and potentially non-equilibrium) process through which coordination takes place as well as the stability
of the eventual equilibrium \cite{Stauffer_06}.
We consider a prototype model to study this kind of situations. In particular, we consider a simple coordination game with $N$ agents and $N$ choices. Individual agents aim to converge to a single universally chosen outcome; i.e., the game can be thought of as a \textit{majority game}.

In the language of game theory, this relates to the idea of equilibrium selection. In our game, there are $N$ possible pure strategy {\it Nash equilibria}, each of which is equally attractive to the agents. The question is how, in the absence of communication, do the agents converge to only one equilibrium? Naturally, we do not allow a central planner to dictate the solution as that would make the problem trivial as well as unrealistic.

In our model, agents play the game repeatedly and they always want to be in the majority. We first present several strategies based on {\it na\"{i}ve learning} that allow the agents to solve this coordination problem in a distributed manner \cite{ChalletJEDC08}.
We next assume that the agents want to minimize their cost of experimentation, i.e., to come up with some fixed strategy as soon as possible even if it results in not being in the absolute majority. This leads to a trade off between the degree of stability (time to attain an approximate fixed rule of thumb) and the efficiency of the solution i.e., degree of coordination.
We propose multiple heuristic strategies for coordination that solves the problem to different degrees. We propose a {\it Polya}
scheme following the famous Polya's urn model, which allows us to interpolate between multiple types of reinforcment learning processes \cite{Sornette2004}.

This paper is intimately related to the literature of minority game \cite{Challet2004,Fogel1999,Ji-QiangZhang;16} and the generalization of the minority game that goes by the name of Kolkata Paise Restaurant (KPR) problem \cite{chakrabarti2009,Ghosh2010}. In the minority game, there are $N$ agents and 2 options to choose from. The agents' objective is to be in the minority. KPR problem extended this to a minority game with $N$ agents and $N$ options labeled restaurants. In spirit of Ref. \cite{Arthur1994a}, multiple attempts were made to propose strategies that uses finite information sets with bounded rationality. Interested readers can refer to Ref. \cite{Chakraborti2015} for a comprehensive review. The model we propose is the exact opposite of the multi-choice minority game. Both are examples of large scale distributed coordination problems that study competing agents employing adaptive strategies with limited learning.
\cite{Challet04}.

In this paper, we show that agents converge to specific choices
due to reinforcement learning. In particular, depending on the degree of reinforcement, agents may be get stuck to different
choices creating clusters of different sizes. 
Clustering behavior has been studied in the context of minority games \cite{Hod_PRL_03}.
Here, such behavior also implies that due to reinforcement, non-equilbrium configurations may also survive and
hence, it is not necessarily a `winner-take-all' scenario.
Finally, we show that if the agents value not only coordination but also the time requirment to achieve absolute coordination, then 
there would be a trade-off in terms of efficiency and stability of the final solution.

\section{$N$-agent coordination game}

We consider $N$ agents and $M$ options. Time is discrete and at every point of time, each agent makes a choice about which among the $M$ options to use. To fix the idea, one can imagine each option to represent one restaurant which an agent will visit in a time slice. Therefore, each of the $N$ agents' strategy is to choose a restaurant to visit in each time slice. Any given restaurant can accommodate a maximum of $N$ agents in any particular time slice.  
The agents' objective is to stay in majority, i.e., the agents would like to move to a restaurant which has higher number of agents.
In principle, $N$ may not be equal to $M$. To impose symmetry on the problem, we assume $N = M$, i.e., the number of agents is equal to the number of restaurants.

We also emphasize here that the game is necessarily non-cooperative and no communication is allowed among the agents. The information set for all agents is constrained to only their history and partial knowledge about past evolution of the restaurant occupancies. Naturally, allowing full set of history across all restaurants to be available to the agents would immediately solve the problem as the agents can employ a strategy that in time slice 1 they choose randomly and in the next time slice, they move to the restaurant that attracted most number of agents in the first time slice. To have a non-trivial solution, we allow only partial set of history to be available to the agents. We elaborate on the specifics of the information sets for each type of strategies below.

\begin{figure}
\includegraphics[width=0.82\linewidth]{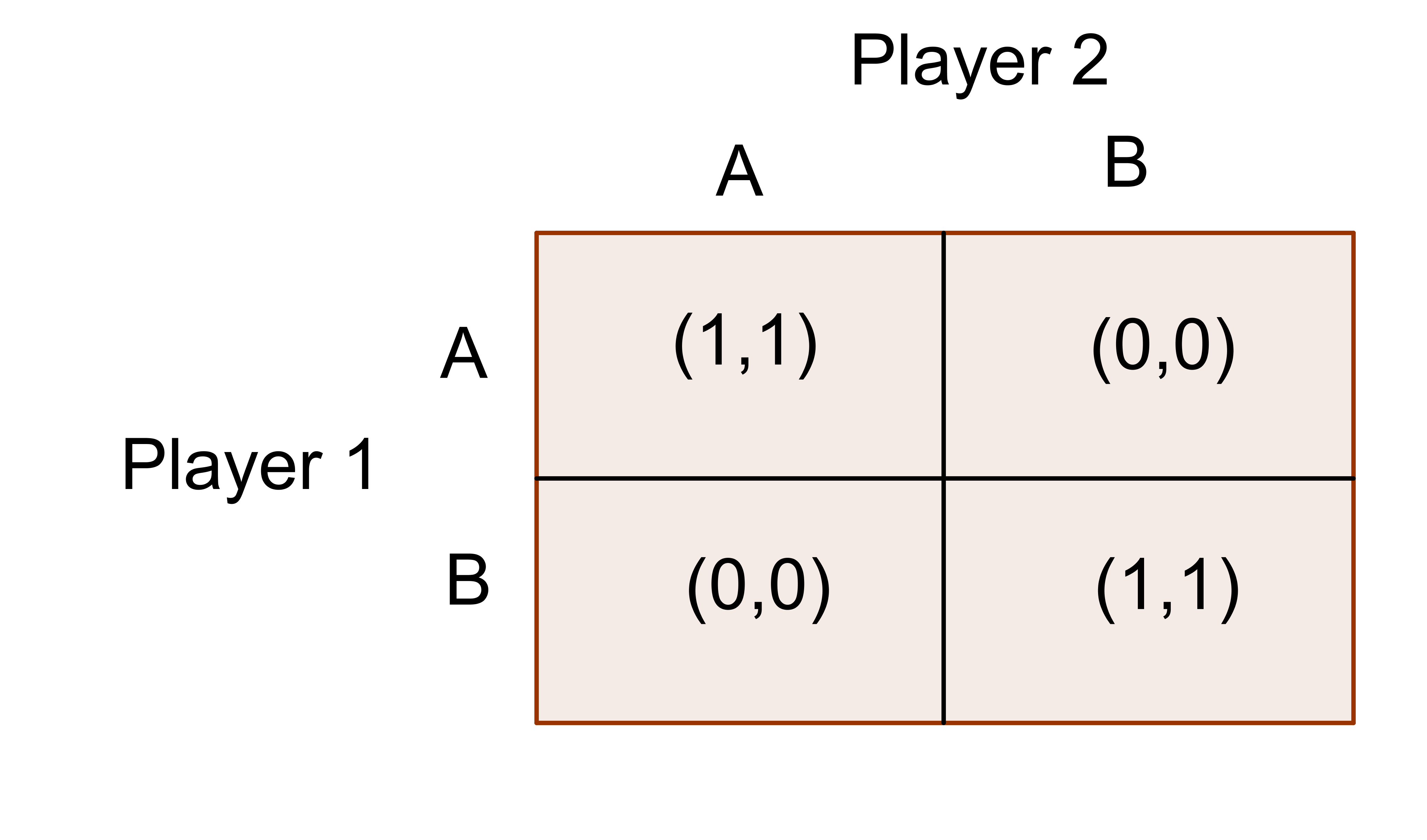}
\caption{Payoff matrix of the coordination game with 2 players. Both A,A and B,B are equilibria.}
\label{fig:payoff_matrix}
\end{figure}

Fig. \ref{fig:payoff_matrix} shows the payoff matrix for a general convergence game for two players. Both players have strategies A and B i.e., they may choose to visit either restaurant A or restaurant B. If both of them decide to visit the same restaurant (either A or B), then the outcome for both would be better than if the chose different restaurants. A couple of points may be noted. This game is a simplified version of the famous {\it Battle of Sexes} game (see for example, \cite{osborne_game_theory} for a textbook treatment). The Battle of Sexes game allows two players, in which agents aim to converge to a single restaurant although they differ in their preferences over the restaurants. In this paper we assume a multi-agent multi-choice scenario with $2 \leq N < \infty$ agents, but assume that all agents have identical preference over the restaurants.

The agents decide on their strategies based on attractiveness of a restaurant. We define attractiveness ($A$) of a restaurant as the number of agents that have chosen that restaurant. Thus attractiveness depends on the information set that the agent possess. Naturally, at any given time slice, it is not possible to know how many other agents 
are choosing a given option.

For the sake of completeness, we define {\it Nash equilibria} for the coordination game. A Nash equilibrium is defined as a strategy collection such that given every other agent's strategy each agent is weakly better off by not switching to a different strategy. For our purpose this description suffices. For a textbook description, see \cite{osborne_game_theory}. From Fig. \ref{fig:payoff_matrix} it can be verified that there are two pure-strategy Nash equilibria, viz. both go to either restaurant $A$ or both go to $B$. In a general $N$-agent game, there would be $N$ pure strategy Nash equilbria.

It may be noted that Nash equilibrium is an  equilibrium description and a static concept. It does not explain how one equilibrium would be chosen from many candidate equilibria in reality. So the essential question is how do agents coordinate to converge on one equilibrium out of $N$ possible choices, in absence of any information about what the other agents are thinking? 

We specify a set of strategies below that solves this problem using finite sets of information and in certain cases, with no information about the other agents.  

\section{Heuristic Updating Strategies}
In this section, we present a set of updating strategies that the agents may employ in the coordination game. These can be thought of as rule-of-thumb strategies. In particular,
they do not exhaust all possible strategies, but provides a comprehensive set that is useful for solving the game.

In the following, we define a strategy of an agent as a vector of probabilities that she assigns to the restaurants i.e. each of the elements of the vector would represent the
probability with which she chooses one restaurant. Formally, we denote the $i$-th agent's strategy at time slice $t$ as $\{p_{ijt}\}$ for $j\in N$.
Learning is introduced as updating the probability vector based on success of failure in the past.

\begin{figure}
\centering
\includegraphics[width=.97\linewidth,trim=10 20 0 25,clip]{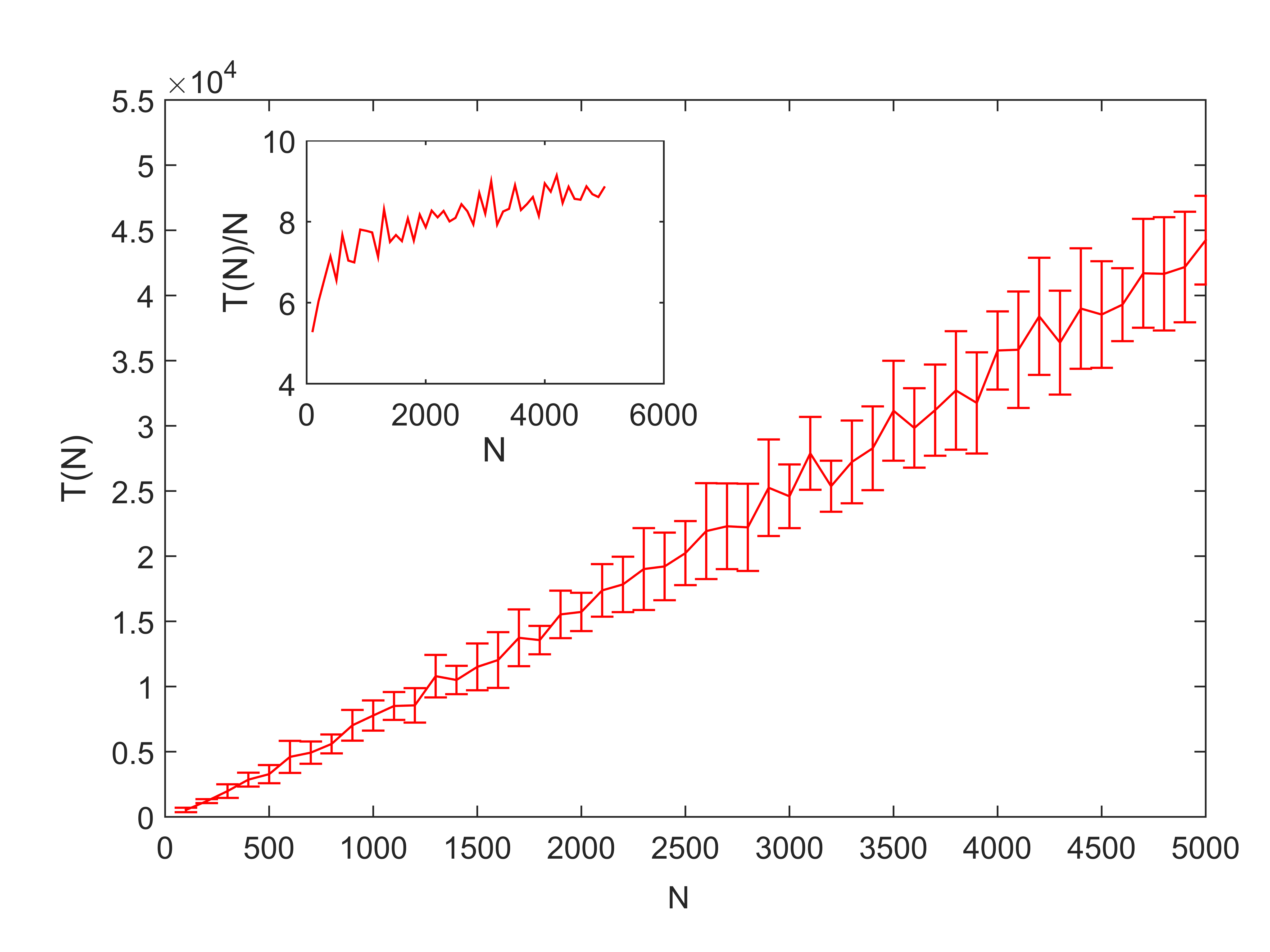}
\caption{Simulation results for the `No learning' strategy. Number of time slices required for convergence, averaged over 10 parallel simulations.$T(N)$ denotes the time of convergence with $N$ number of agents. The vertical bars shows standard deviation of the of simulation results. In the inset, we plot $T(N)/N$ as a function of the system size $N$, which stabilizes around 8.5. 
Thus time required for convergence scales linearly with $N$.}
\label{fig:RangeofmultipleofNforconvergence}
\end{figure}

\begin{figure}
\centering
\includegraphics[width=0.97\linewidth,trim=10 25 0 25]{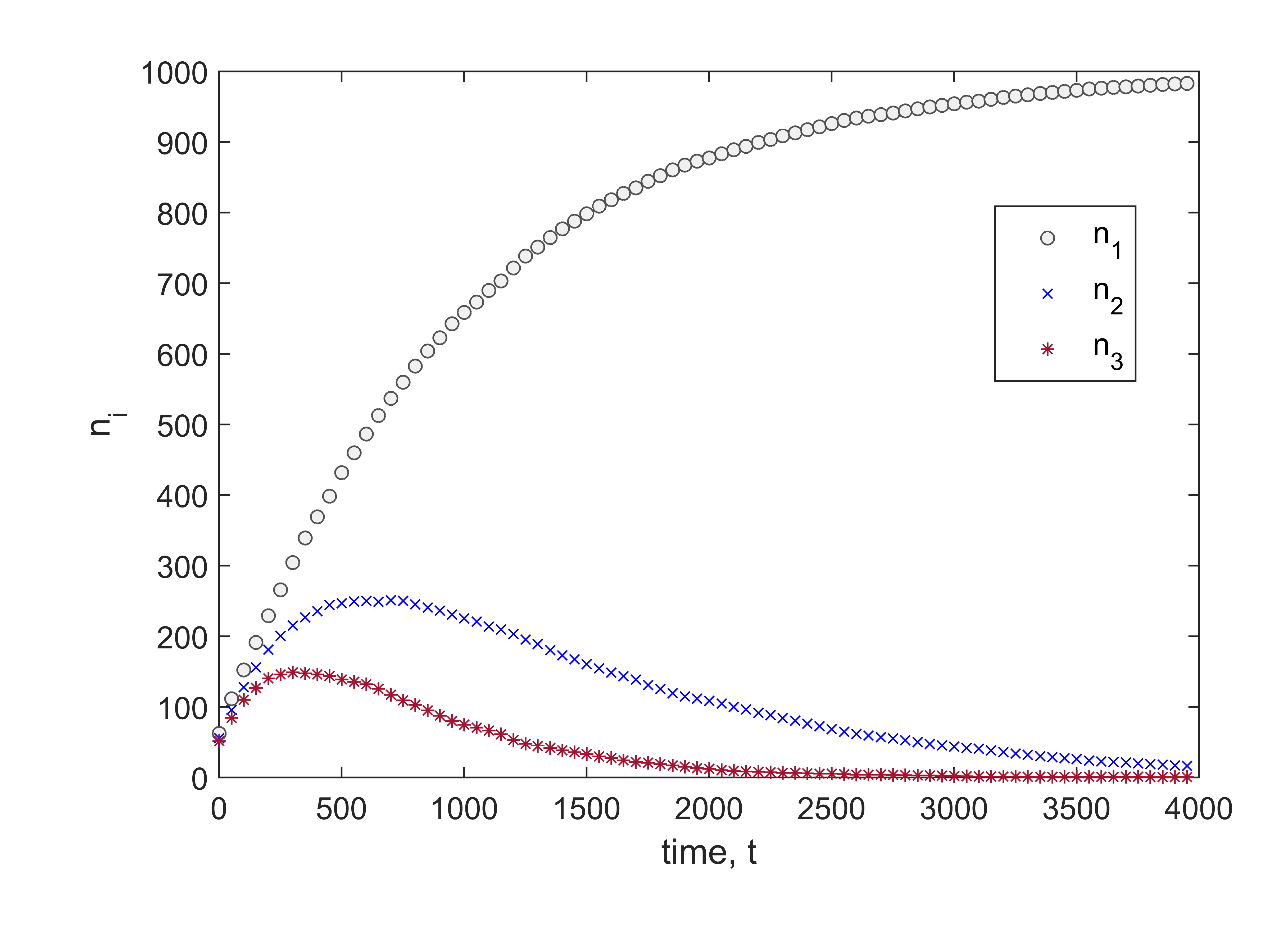}
\caption{Simulation results for the `No learning' strategy. This strategy leads to convergence linearly with time. On $y$-axis  we plot the number of people in the restaurant with largest  (red), 2nd largest (black) and 3rd largest (blue) no. of agents. On the $x$-axis we plot time.}
\label{fig:Evolution_of_maximum_number_of_agents_in_any_option}
\end{figure}

\subsection{No learning} 
We begin with a {\it No Learning} strategy. This entails zero probability updating and represents a baseline case.

\subsubsection{Zero updating}
This strategy has two parts. Consider any generic time slice $t$. First,
the $i$-th agent ($i\in N$) assigns the following probability to the restaurants,
\begin{equation}
p_{ijt}=\frac{1}{N}.
\label{Eqn:no_learning}
\end{equation}
Naturally, this would lead to a randomly distributed allocation of agents across restaurants. In particular, \cite{KPR} shows that the occupancy fraction i.e. the number of
restaurants occupied as a fraction of the total number $N$, would be 63.5\%. So the first part is far from sufficient to ensure coordination.

The second part of the strategy allows the agent at time slice $t$, to make a comparison between the choice made at time slice $t$ and the restaurant she is at time slice $t$.
Because attractiveness depends on the number of agents in a restaurant, we denote the $j$-th restaurant's attractiveness at time slice $t$ by $A_{jt}$. 
Therefore, an agent's strategy who is at restaurant $k$ is to go to restaurant $j$ if
\begin{equation}
A_{jt}\ge A_{kt},
\end{equation}
else, the agent stays at $k$.

\vskip .5 cm

\paragraph{Information required:}  The information set of the $i$-th agent who is at restaurant $K$ at the $t$-th time slice comprises $A_{kt}$ and $A_{jt}$ where 
$j$ is the outcome of random selection scheme (Eqn. \ref{Eqn:no_learning}) for the $i$-th agent. Note that it entails gathering information about the $j$-th restaurant that
the $i$-th agent has not visited at time slice $t$, implying that we are allowing for local information. In principle, one can imagine that the agents may have to pay a cost to gather that information. This is a point we will later take on in fuller details.		
	
\vskip 0.5 cm

\paragraph{Results:}  We present simulation results in Fig. \ref{fig:RangeofmultipleofNforconvergence} and Fig. \ref{fig:Evolution_of_maximum_number_of_agents_in_any_option}. Fig. \ref{fig:RangeofmultipleofNforconvergence} shows the time required for absolute convergence $T(N)$ i.e. the minimum number of time slices required for all agents to converge at one restaurant, as a function of the number of agents $N$. It shows a linear trend with a coefficient about 8 on an average. In the inset, we show the ratio $T(N)/N$ as a function of $N$ which fluctuates around 8 after an initial steep rise. In the main diagram, we also provide an estimation of the standard deviation across $O(10)$ number of simulations.

Fig. \ref{fig:Evolution_of_maximum_number_of_agents_in_any_option} shows the dominance of one restaurant over others (we show the second and the third most populated ones) over time in one simulation with $N=1000$. The second and the third most crowded restaurant initially starts attracting more agents before decaying completely in terms of the number of agents as the dominant one becomes absolutely dominant and attracts all agents.

These results show that symmetry-breaking occurs due to stochastic choices. All restaurants start off by being equally popular. But at the end, only one of them emerges as the most popular choice
and all other restaurants have no agents. 

\subsection{Learning Strategies} 
In this section, we introduce updating rules based on success and failures of the past choices.

\subsection{Ex-ante knowledge} 
\label{Subsec:ex-ante}
This is a direct extension of the previous strategy. At each time slice, the $i-$th agent ($i\in N$) makes a choice of restaurants using a probability vector $\{p_{ijt}\}~\forall~j\in N$. Then she compares the attractivenesses of the chosen restaurant and the restaurant she is currently in, and moves to the one with higher attractiveness in the next time slice. Finally, the $i$-th agent updates her probability vector based on the attractiveness. This last step of probability updating differentiates the strategy from the No Learning strategy.

We call this strategy \textit{ex-ante} as the agents can decide whether or not to move to a chosen restaurant by gathering information about attractiveness of the current restaurant and the newly chosen one. 
Later in Sec. \ref{Subsec:ex-post} we study a case with ex-post  updating that relaxes this assumption.

We extend the strategy under consideration in multiple dimensions. 
In the first case, agents reward for  higher attractiveness and punishment for lower attractiveness. Formally, higher attractiveness implies that the agent would assign higher weight
in the probability vector and would reduce weight for restaurants with lower attractiveness. This strategy we label as {\it symmetric} in updating.

In the second case, the agents only reward higher attractiveness. We label this strategy as {\it asymmetric} updating.
Further, we consider the cases where the agents are allowed to choose more than one restaurant to pick the best option.
Formally, the information set increases to $k$ choices per agent, where $k=1,2,3,\ldots$ etc. Naturally, setting $k=N$ makes the problem trivial. So we concentrate on cases with sufficiently small values of $k$.

Below we describe the strategies in details.

\subsubsection{Symmetric updating}
\label{Subsubsec:ex-ante_sym}

Consider agent $i$ where $i\in N$, at any generic time slice $t$. Suppose she is at restaurant $r$ and given her probability vector $\{p_{ijt}\}$, she probabilistically picks restaurant $l$.
If $A_{lt}<A_{rt}$, she stays at restaurant $r$. Else, she moves to restaurant $l$.

Simultaneously, the agent updates probability of restaurants $l$ and $r$ such that the one with higher attractiveness will gain in probability by a fraction ($f_1$) while the other  will decrease by fraction ($f_2$). Naturally, the resulting sum is normalized to 1. Formally,
if $A_{lt}<A_{rt}$,
\begin{align*}
p_{ij(t+\frac{1}{2})}	 =
\begin{cases} 
p_{ijt}+f_1(1-p_{ijt})  & \text{ for $j=r$,} \\
p_{ijt}-f_2(p_{ijt})  & \text{ for $j=l$.} \\
\end{cases} 
\end{align*}
If $A_{lt}=A_{rt}$,
\begin{equation*}
p_{ij(t+\frac{1}{2})}	 = p_{ijt}\quad\text{for $j\in N$,}  
\end{equation*}
and if $A_{lt}>A_{rt}$,
\begin{equation*}
p_{ij(t+\frac{1}{2})}	 =
\begin{cases} 
p_{ijt}+f_1(1-p_{ijt})  & \text{for $j=l$,} \\
 p_{ijt}-f_2(p_{ijt})  & \text{for $j=r$.} \\
\end{cases}  
\end{equation*}
Finally, probabilities are normalized:
\begin{equation*}
p_{ij(t+1)}=p_{ij(t+\frac{1}{2})}/\sum_i p_{ij(t+\frac{1}{2})}.
\end{equation*}

%

\paragraph{Information required:}   The information set is identical to the No Learning strategy for $k=1$. For higher values of $k$, we allow the agents to have more information about the occupancy of the restaurants in the previous time slice to make a comparison.

\begin{figure}
	\centering
	\includegraphics[width=0.95\linewidth,trim= 10 20 0 10,clip]{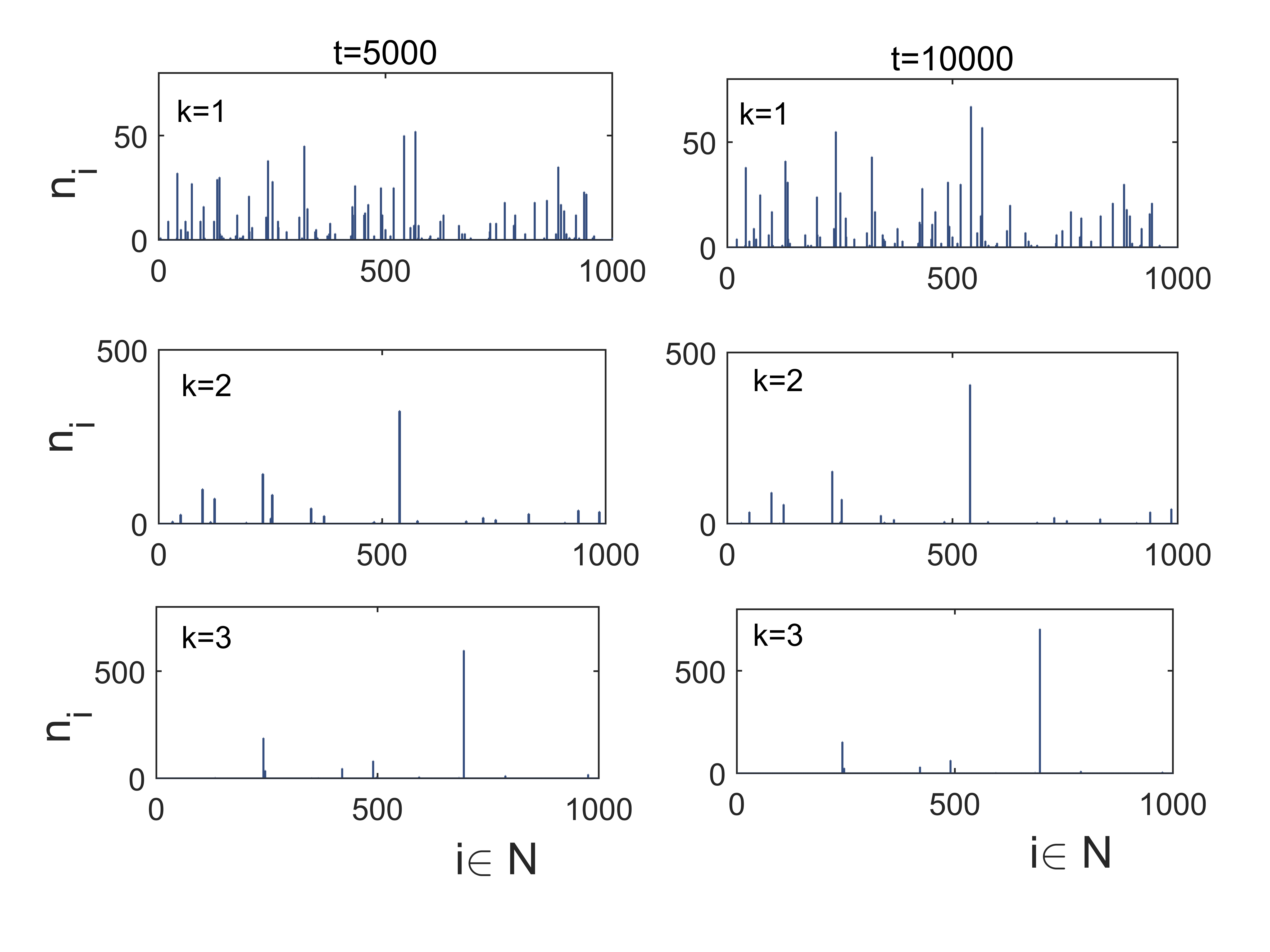}
	\caption{Simulation results for `ex-ante updating' strategy with symmetric reinforcements (success is rewarded ($f_1=1)$ and failure is punished ($f_2=0.1)$). We present two snapshots (left column at $t=5000$ and right column at $t = 10000$) of possible evolutions of the system with $N=1000$ agents. The rows show the results for different values of the information sets, $k=1$, $k=2$ and $k=3$. As is evident, with increasing size of the 
		information set, convergence occurs faster.}
	\label{fig:exante_s_1}
\end{figure}	

\paragraph{Results:}   Fig. \ref{fig:exante_s_1} shows the simulation results for this strategy wih $N=1000$.
On the $x$-axis, we plot the restaurants and on the $y$-axis, we plot the number of agents that goes to the restaurants $n_i$ for all restaurants i.e. for all $i\in N$.
We show two snapshots. One at time slice $t=$ 5000 and the other at $t=$ 10000. The three rows show the distribution of agents under three different information sets, $k=1,2,3$.

The first thing to notice is that the dynamics becomes considerably slow. Even after 10000 time slices (for $N=1000$), attaining coordination is very difficult as the panels on the right in Fig. \ref{fig:exante_s_1} show very clearly. However, we note that convergence is guaranteed. As an explanation, consider a case with distribution of agents across all restaurants as  (501,499,0,....,0). Given the current strategy, only the first restaurant can attract agents and the second one can only lose agents however slow the process might be.

The next important feature is that by increasing the information set even by limited amount (going from $k=1$ to 2 and 3) drastically improves degree of coordination although
the dynamics becomes slow after a certain  point. For example, in the bottom row, we see that the distribution changes very slowly going from  $t= 5000$ to $t= 10000$.

Therefore, we see that for a long time there are clusters of agents in different restaurants before all collapse into one giant cluster i.e. absolute convergence takes place.
Such clustering behavior is transitory. 

\subsubsection{Asymmetric updating}
\label{Subsubsec:ex-ante_asym}

Consider agent $i$ at time $t$ in restaurant $r$, probabilistically picking another restaurant $l$. If $A_{lt}<A_{rt}$, she stays at restaurant $r$. Else, she moves to restaurant $l$.
The asymmetric updating scheme differs from the symmetric scheme in the way she updates the probability vector $\{p_{ijt}\}$.

If there is a difference between attractiveness of the current restaurant and the probabilistically picked one, the agent assigns a higher weight to the more attractive option and reduce weight for every other restaurants.
Formally, if $A_{lt}<A_{rt}$ 
\begin{align*}
p_{ij(t+\frac{1}{2})}	 =
\begin{cases} 
p_{ijt}+f(1-p_{ijt}) & \text{ if $j=r$} \\
(1-f)p_{ijt} & \text{otherwise.} \\
\end{cases}
\end{align*} 
If $A_{lt}=A_{rt}$,
\begin{equation*} 
p_{ij(t+\frac{1}{2})}	 = p_{ijt}\quad\text{for $j\in N$},
\end{equation*}
and if $A_{lt}>A_{rt}$
\begin{align*}
p_{ij(t+\frac{1}{2})}	 =
\begin{cases} 
p_{ijt}+f(1-p_{ijt})  & \text{ if $j=l$} \\
(1-f)p_{ijt} & \text{otherwise.} \\
\end{cases}  
\end{align*}
Finally, probabilities are normalized:
\begin{equation*}
p_{ij(t+1)}=p_{ij(t+\frac{1}{2})}/\sum_i p_{ij(t+\frac{1}{2})}.
\end{equation*}

%

\paragraph{Information required:}   This strategy requires exactly the same set of information as the {\it symmetric} updating strategy.	

\begin{figure}
	\centering
	\includegraphics[width=0.95\linewidth,trim=20 10 15 10,clip]{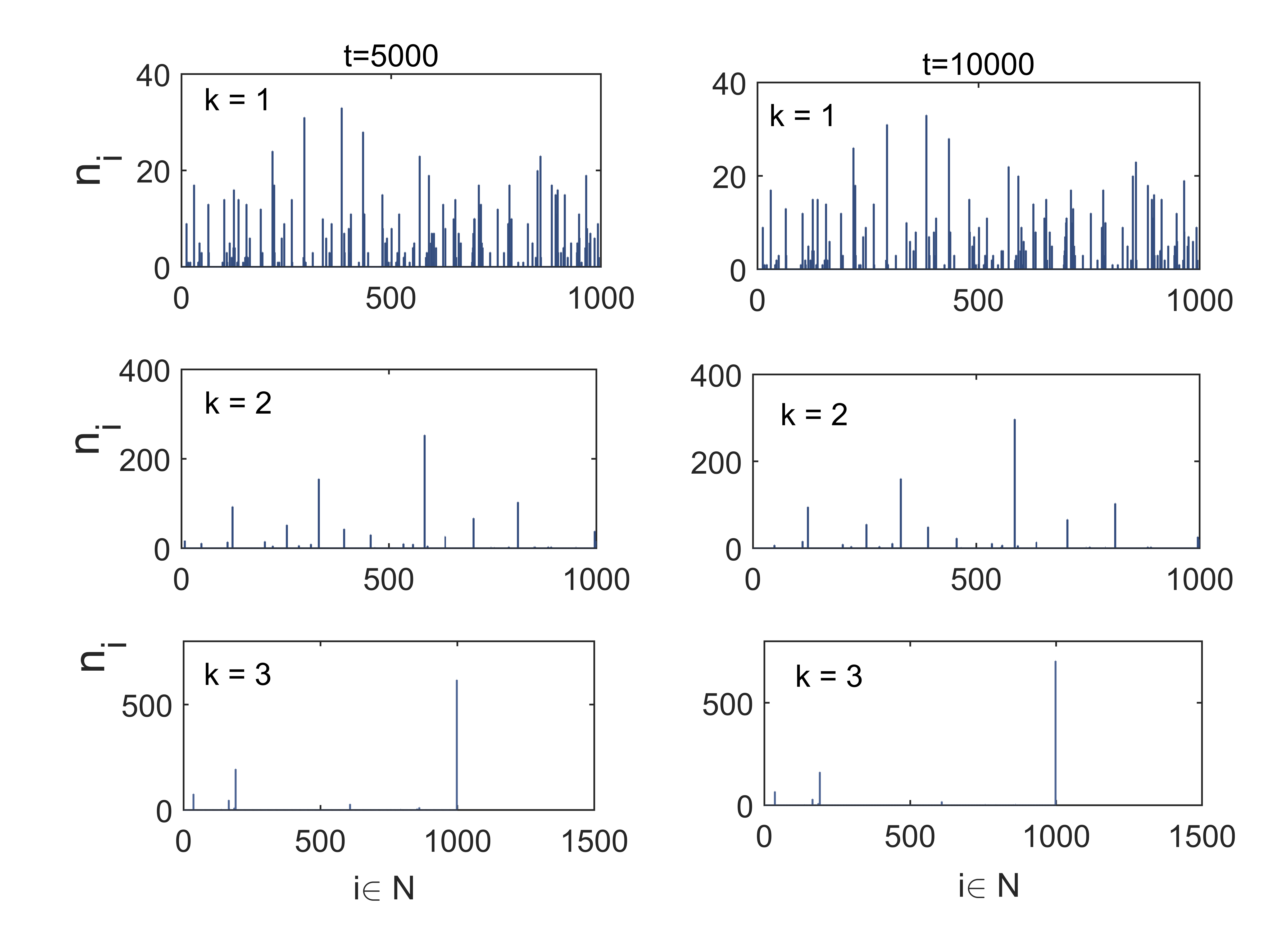}
	\caption{Simulation results for `ex-ante updating' strategy with asymmetric reinforcements where only success is rewarded ($f=0.25$). We present two snapshots (left column at $t=5000$ and right column at $t =  10000$) of possible evolutions of the system with $N=1000$ agents. The rows show the results for different values of the information sets, $k=1$, $k=2$ and $k=3$. As is evident, with increasing size of the 
		information set, convergence occurs faster as was the case with the symmetric updating rule.}
	\label{fig:exante_s_2}
\end{figure}

\paragraph{Results:}   Fig. \ref{fig:exante_s_2} presents the simulation results with the asymmetric updating strategy for $f=0.25$. The results are comparable to the
{\it symmetric} updating scheme. We see that the dynamics becomes slow. As we expand the information set from $k=1$ to 2 and 3, convergence takes place much faster in the initial phase. But after
a while, it becomes slow for all information sets. But again with sufficient number of iterations, absolute convergence takes place.

\begin{figure}
	\centering
	\includegraphics[width=0.95\linewidth,trim=20 20 15 10,clip]{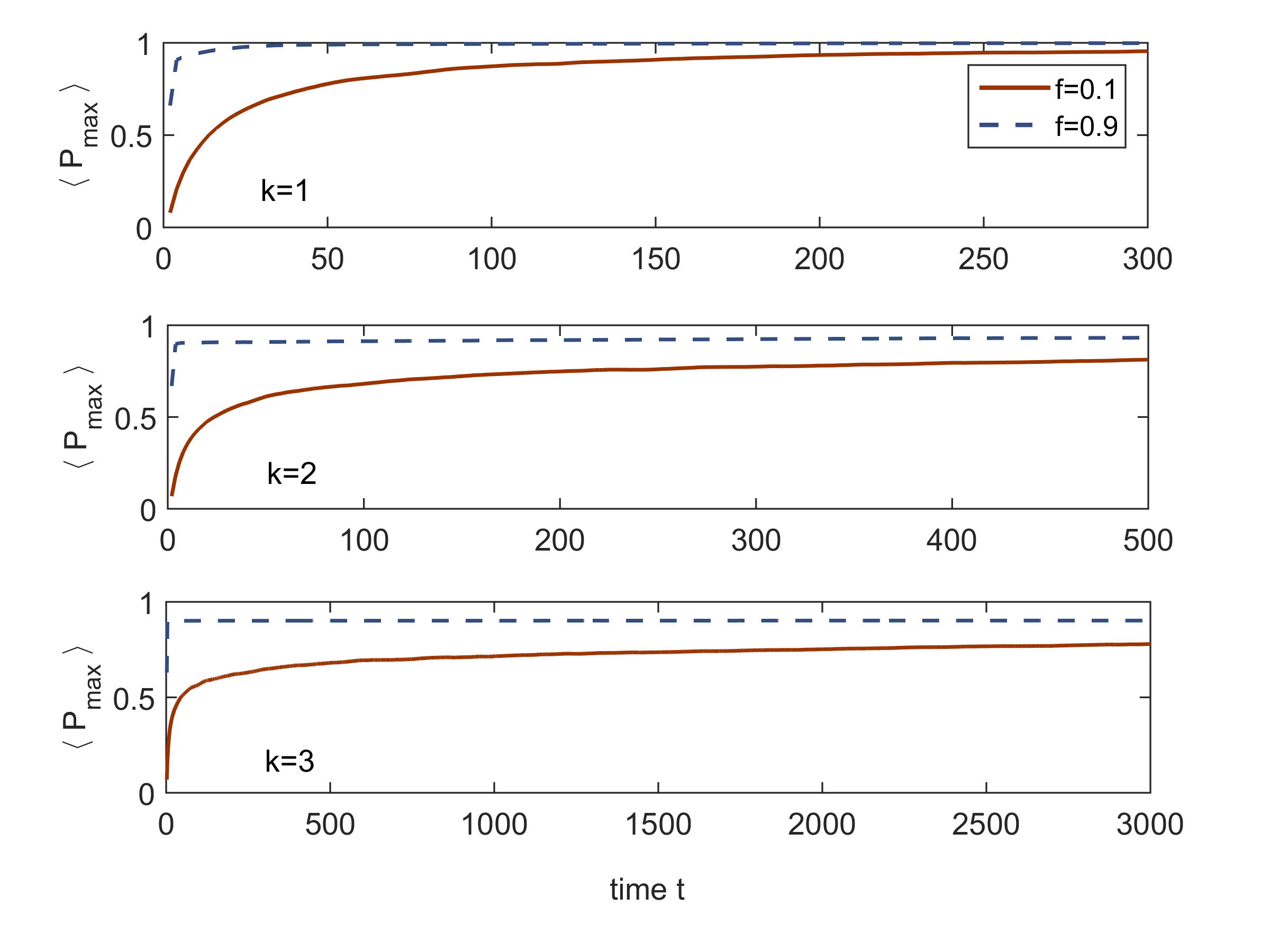}
	\caption{Simulation results for `ex-ante updating' strategy with asymmetric reinforcements with $f$ = 0.1 and 0.9 and three information sets, $k=1, 2$ and 3. 
		We show the evolution of the average of the maximum probability that the agents assign to any one restaurant at all time slices. With high reinforcement ($f=0.9$), the maximum probability converges much faster than with low reinforcement ($f=0.1$).}
	\label{fig:f_analysis_k_123}
\end{figure}
Therefore, we see that for a long time there are clusters. But as with the {\it symmetric } updating, this behavior is transitory. By varying the parameter $f$
we studied the dynamics before convergence. Fig. \ref{fig:f_analysis_k_123} presents simulation results for two different values of $f$ with multiple information sets ($k=1,2,3$).
in order to quantify the degree of stability before convergence, we compute the average of the maximum probabilities that the agents assign to any restaurant.
With smaller values of $f$ ($f$ = 0.1), the average probability goes up very fast compared to larger values ($f=0.9$). 
Also, with bigger information sets, the average of the maximum probabilities rise slower than with smaller information sets.
This is consistent with the finding that coordination occurs much faster with bigger information sets, as that requires multiple switching to ensure
convergence. Naturally, with switching happening at a higher frequency leads to lesser reinforcements to specific restaurants.

\subsection{Reinforcement learning through Polya's urn model}

We introduce a new strategy using the Poly's urn model (\cite{Sornette2004}) that effectively captures reinforcement learning.
Let us define 
\begin{equation}
\phi=\frac{mn}{N-m} 
\label{Eqn:phi_Polya}
\end{equation}
where $m$ is a tunable parameter taking discrete values within 0 and $N$. We denote the number of times the $i$-th agent has visited restaurant
$j$ before time slice $t$, by $n_{ijt}$. Then the probability of choosing restaurant $j$ is given by
\begin{equation}
p_{ijt}=\frac{1+\phi n_{ijt}}{N+\phi \sum n_{ijt}}.
\end{equation}
Intuitively, this is
an extension of the basic {\it No Learning} strategy (which would require $p_{ijt}=1/N$) by embedding reinforcement learning through Polya's urn model.

\paragraph{Information required:}   The required information set for the $i$-th agent is derived only from the full sequence of success of the agent at different restaurants.
It is reasonable to assume that the agents keep track of their own visits. Also note that at any time slice, the agent does not require any information from a restaurant that she is not visiting as was required with the earlier strategies. This is possible because there is no comparison involved. The probabilistic strategies are devised based on historical success.

\begin{figure}[hbt]
	\centering
	\includegraphics[width=0.95\linewidth,trim=15 5 15 10,clip]{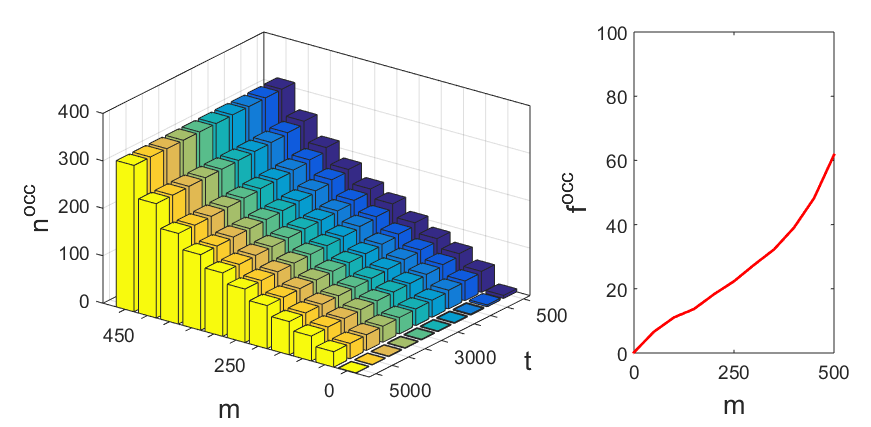}
	\caption{
		{\it Left panel}: Evolution of the number of restaurants ($n^{occ}$) occupied as function of Polya factor $m$ from 0 to 495 for $N$=500, $t$=5000. In the limit $m\rightarrow N$, there is infinite reinforcement.
		We can analytically show that the occupancy ratio in that case would be 63.2\%. In the other limit $m\rightarrow 0$, it converges to the {\it No Learning} strategy and hence
		majority problem is solved in linear time as we have shown in Fig. \ref{fig:RangeofmultipleofNforconvergence}.
		{\it Right panel}: Evolution of fraction of restaurants ($f^{occ}$) occupied as function of Polya factor $m$ from 5 to 495 for $N$ = 500, $t$ = 500 to 5000. As is evident, for small $m$ the number of restaurants occupied is very small and in the other extreme, the occupancy ratio is close to 63.2\% ($\approx$ 318/500) which corroborates earlier results.
	}
	\label{fig:occupancy_analysis_3d}
\end{figure}

\paragraph{Results:}   Fig. \ref{fig:occupancy_analysis_3d} presents numerical results for different values of $m$ (see Eqn. \ref{Eqn:phi_Polya}) with $N$ = 500
and $t$ = 5000. In the left panel, we show the number of restaurants occupied ($n^{occ}$) with at least one agent for different values of the factor $m$ and
at different time slices $t$. Clearly when $m=0$, the Polya's scheme would converge to {\it No Learning} case and absolute convergence occurs. This implies
only one restaurant would be occupied. This can be seen from the figure by looking at the bars for different time slices by fixing $m$ = 0. In the other extreme with $m=N-5$
(for simulations, we can not set $m=N$), 
we see that around 318 restaurants out of 500 have been occupied. This is consistent with the notion that setting the factor $m$ very close to $N$
leads to infinite reinforcement implying if an agent goes to one restaurant, she would stick there for the rest of the time slices. So effectively,
the choices in the first time slice itself determines the distribution of agents across restaurants as that distribution will never change because of infinite reinforcement.
It is easy to show that as the agents are starting with uniformly distributed probabilities ($p_{iuj0}=1/N$), in the first time slice 63.2\% of the restaurants would be occupied.
We are skipping the derivation of this fraction. Interested readers can refer to \cite{KPR}.
One can easily verify that 318/500 is close to 63.5\% and hence this validates our results.
The right panel in the same figure shows the fraction of restaurants occupied i.e. $f^{occ}=n^{occ}/N$. The results are perfectly consistent with the left panel.

We also note that having $m=N$ in {\it Polya's scheme} (i.e. infinite reinforcement) is identical to assuming 
$f=1$ in the {\it asymmetric} updating strategy. Thus in the limit, these two strategies are exactly identical.
This strategy allows us to interpolate between a wide spectrum of reinforcement by changing the factor $m$.
In particular, it allows us to cover the same range as are separately done by the {\it symmetric} and {\it asymmetric} updating strategies.

\subsection{Ex-post knowledge} 
\label{Subsec:ex-post}

In the case of {\it ex-ante knowledge} in Sec. \ref{Subsec:ex-ante}, we studied strategies where the agents can obtain information about the newly chosen restaurant's
attractiveness and make a comparison between the chosen restaurant's and the current restaurant's attractiveness. However, this might be a costly activity
to know the attractiveness of  another restaurant before actually visiting it. In the present section, we study the same set of strategies where the agents can obtain
information about attractiveness only after she moves to the chosen restaurant. An important distinction from the earlier cases is that the present strategy
allows for {\it regret}. After the agent moves to a new restaurant, she comes to know about its attractiveness and hence cannot do comparison prior to switching.
Updating the probability vector happens the same way depending on relative attractiveness as was done in Sec. \ref{Subsec:ex-ante}.

\subsubsection{Symmetric updating}
\label{Subsubsec:ex-post_sym} 
Consider agent $i$ where $i\in N$, at any generic time slice $t$. Suppose she is at restaurant $r$ and given her probability vector $\{p_{ijt}\}$, she probabilistically picks restaurant $l$.
After knowing both $A_{lt}$ and $A_{rt}$, probability vector $p_{ijt}$ is updated exactly the same way as in Sec. \ref{Subsubsec:ex-ante_sym}.
To avoid repetition, we are skipping the probability updating schemes.

\paragraph{Information required:}  The required information comes from the restaurants that the agent has visited. Hence, there is no external information acquired.

\begin{figure}[hbt]
\centering
\includegraphics[width=0.95\linewidth,trim=10 10 15 10,clip]{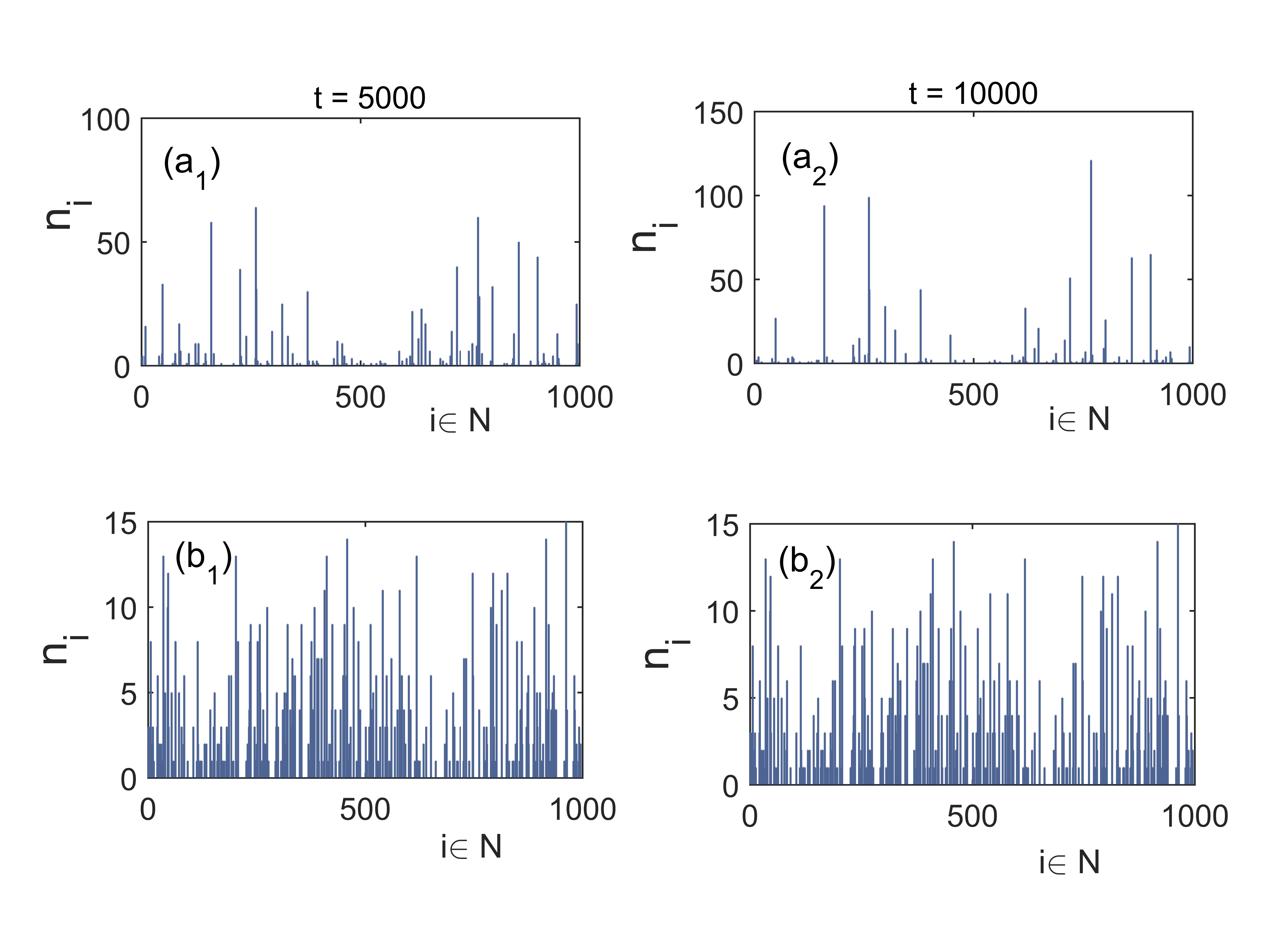}
\caption{Simulation results for `ex-post updating' strategy with symmetric (top panels, $f_1=1$ and $f_2=0.1$) and asymmetric (bottom panels, $f=0.5$) reinforcements. On the $x$-axis, we plot the identity of the restaurants. Symmetric updating leads to higher crowding for the restaurants.}
\label{fig:expost_s_1_s_2}
\end{figure}

\paragraph{Results:}  Fig \ref{fig:expost_s_1_s_2} shows the simulation results in the top panels for $f_1=1$ and $f_2=0.1$. We show results fr two time slices, at $t= 5000$ and
$t=10000$. As in the earlier case, this strategy is also quite slow but eventually converges to a single restaurant in the limit. Naturally, this is slower than the {\it ex-ante knowledge}
case.

 \subsubsection{Asymmetric updating}
 \label{Subsubsec:ex-post_asym}
Similar to above, consider agent $i\in N$, at any generic time slice $t$. Suppose she is at restaurant $r$ and given her probability vector $\{p_{ijt}\}$, she probabilistically picks restaurant $l$.
After knowing both $A_{lt}$ and $A_{rt}$, probability vector $p_{ijt}$ is updated exactly the same way as in Sec. \ref{Subsubsec:ex-ante_asym}.

\paragraph{Information required:}   Required information comes solely form the restaurants she visited and hence no external information is acquired.	

\begin{figure}[hbt]
\centering
\includegraphics[width=0.95\linewidth,trim=20 20 15 10,clip]{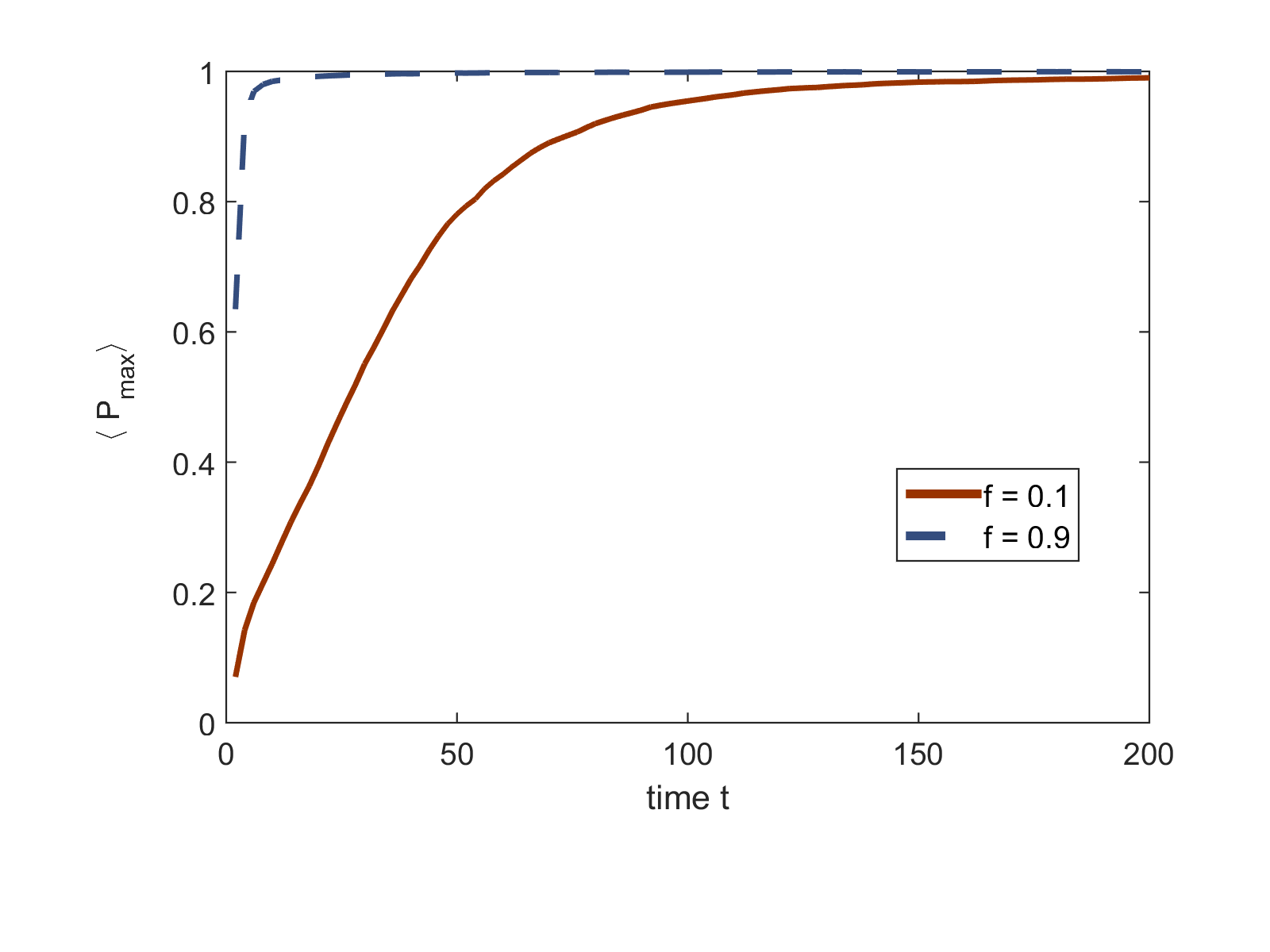}
\caption{ 
Simulation results for `ex-post knowledge' strategy with asymmetric reinforcements with $f$ = 0.1 and 0.9 for a system size $N=1000$.
We show the evolution of the average of the maximum probability that the agents assign to any one restaurant at all time slices. With high reinforcement ($f=0.9$), the maximum probability converges much faster than with low reinforcement ($f=0.1$), similar to the case of 'ex-ante updating' strategies. }
\label{fig:Fanaysis0-250}
\end{figure}

\paragraph{Results:}   Simulation results have been reported in the lower panels of Fig. \ref{fig:expost_s_1_s_2} and Fig. \ref{fig:Fanaysis0-250}. We see that the result for distribution of agents across the restaurants are qualitatively similar to those in the case of {\it symmetric} updating except that coordination is poorer as there are many
restaurants with small numbers of agents.

We study the degree of stability of the transient clusters thus formed in Fig. \ref{fig:Fanaysis0-250}. For higher values of $f$, the average over maximum probabilities
rises quite fast compared to lower values of $f$ though eventually their behavior is similar.
\section{Self-organization and coordination}

In the present section, we discuss the extent to which self-organization occurs in the multi-agent system that solves the coordination problem.

\subsection{Emergence of coordination}

We have seen that some of the strategies especially those which require ex-ante information or knowledge can in principle be thought of as requiring some costs to be payed in order to acquire the information. Also realistically, the agents might have a trade-off in terms of how quickly they can converge to a solution versus the efficiency of the solution. That is they may find it useful to be in majority, not necessarily absolute majority, at a lower time to reach the solution.

A parallel theme is that initially all restaurants are identical. But with absolute convergence, only one of them emerge as the winner. This can be interpreted as how a specific social norm may emerge from multiple possibilities that are a priori equally likely.

Thus emergence of absolute coordination has two contributing factors that can be potentially costly. The first one is obviously the cost of lack of coordination. The second one is the cost of waiting to reach coordination. This can be most clearly seen in the clustering behavior where multiple choices survive as the agents achieve partial coordination reasonably fast.

\begin{figure}
\centering
\includegraphics[width=0.95\linewidth,trim=20 10 15 10,clip]{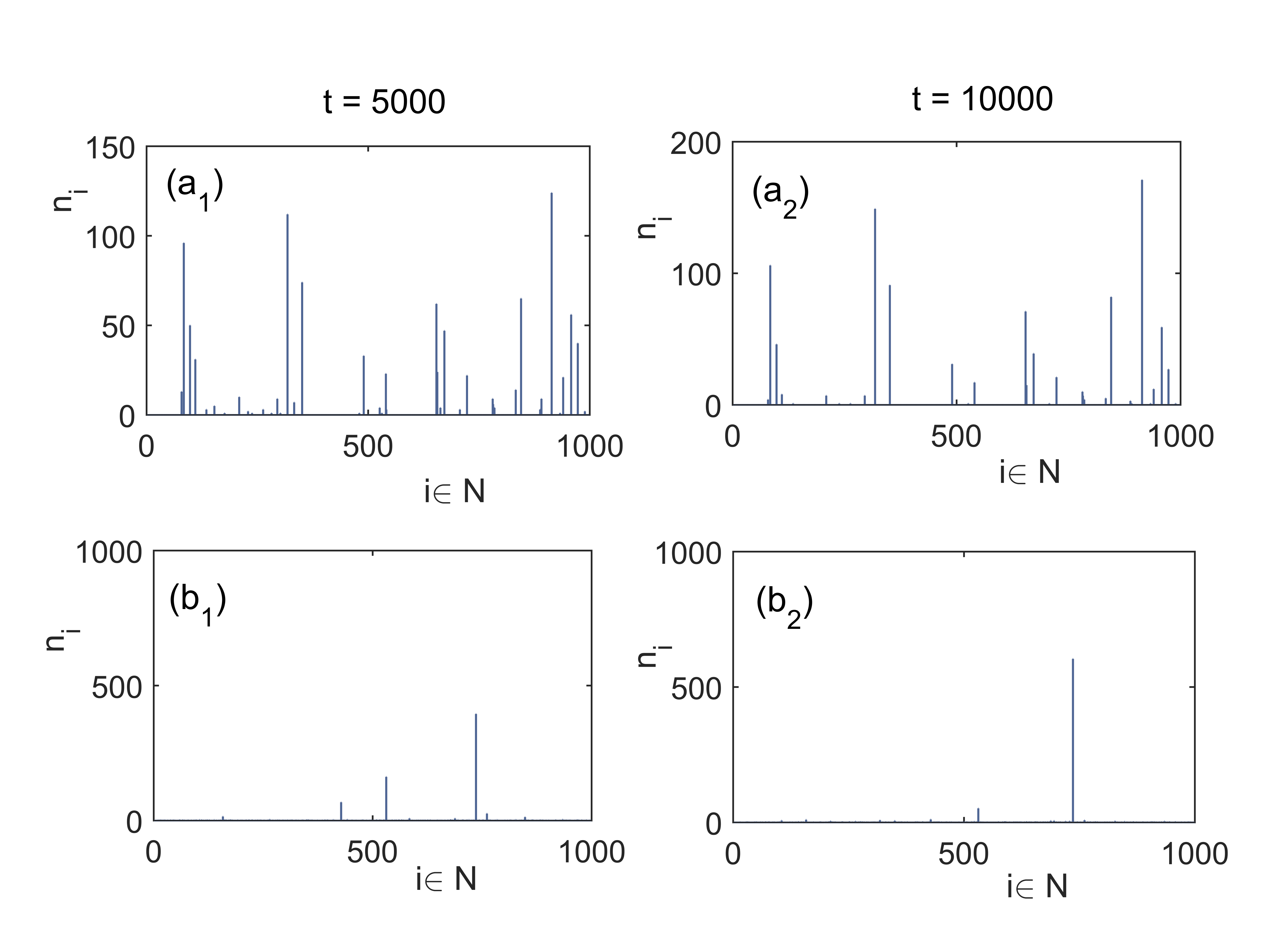}
\caption{ 
Simulation results for `ex-ante' and `ex-post' strategies with asymmetric reinforcements with $f$ = 0.5 and 0.5 for a system size $N=1000$.
Top panels show evolution of coordinaiton across agents with `ex-ante knowldge' and bottom panels show the same with `ex-post knowledge'.
}
\label{fig:compare_exante_expost}
\end{figure}

\subsection{Cluster formation}

We have already seen that clustering behavior can be transient but in almost all cases they are very slowly evolving. This implies that we observe clusters of agents in different restaurants for a very long time. 
\begin{figure}
	\centering
	\includegraphics[width=0.95\linewidth,trim=7 10 15 10,clip]{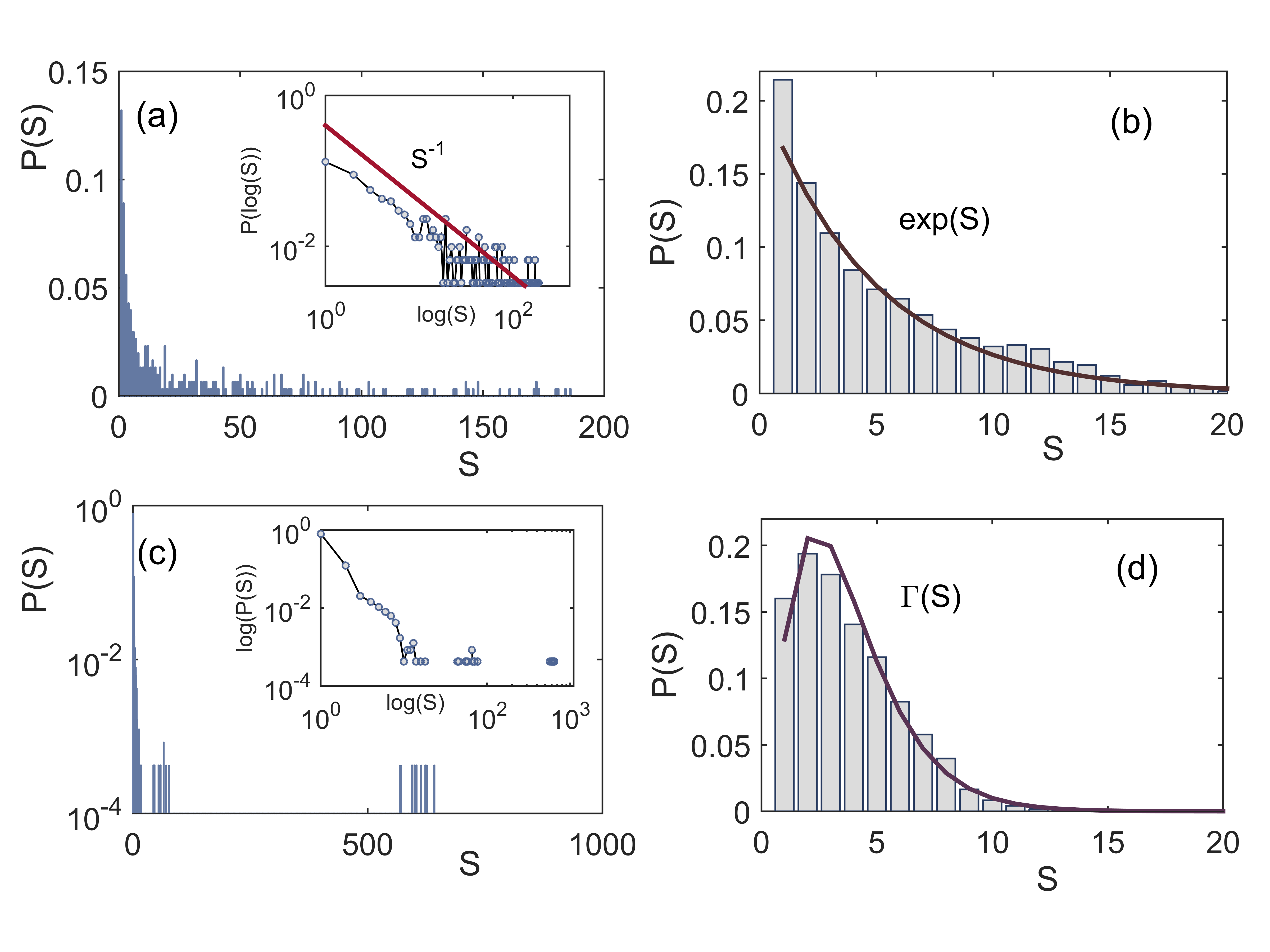}
	\caption{Size distribution of clusters  ($N$ = 1000, $T$ = $10^4$, averaged over $O(10)$ parallel simluations). Panel (a): 
Ex-ante knowledge with symmteric updating (inset: power law fit in log-log plot), panel (b): 
Ex-ante knowledge with asymmteric updating, fitted with exponential distibution, Panel (c): 
Ex-post knowledge with symmteric updating (inset: log-log plot shows the discontinuity in the distribution),
Panel (d): 
Ex-post knowledge with asymmteric updating (inset: fitted with a gamma distribution).
}
	\label{fig:cluster}
\end{figure}
Fig. \ref{fig:cluster} shows four instances of probability density function of clusters. 
We tracked choices of $N=1000$ agents over $t = 10,000$ time slices. We assumed all four cases (ex-ante, ex-post and symmetric-asymmetric) with the previosly mentioned parameter values.
The resulting probability density function has been averaged over $O(10)$ number of simulations. 
Both ex-ante and ex-post with symmetric updating rules (panels (a) and (c)) show strong clustring behavior whereas the other two cases show very moderately distributed clusters (panel (b): fit with exponential distribution with paramter value 4.8564; panel (d): fit with gamma distribution with paramter values 2.7103, 1.3834).

\subsection{Efficiency and cost of waiting}
As discussed before, the agents may have a cost to execute the strategies and hence if there are strategies that takes very long time to reach a state of
absolute convergence, the agents may prefer less efficient solution i.e. smaller clusters, if that is achievable soon enough.
\begin{figure}[hbt]
\centering
\includegraphics[width=0.95\linewidth,trim=20 10 15 10,clip]{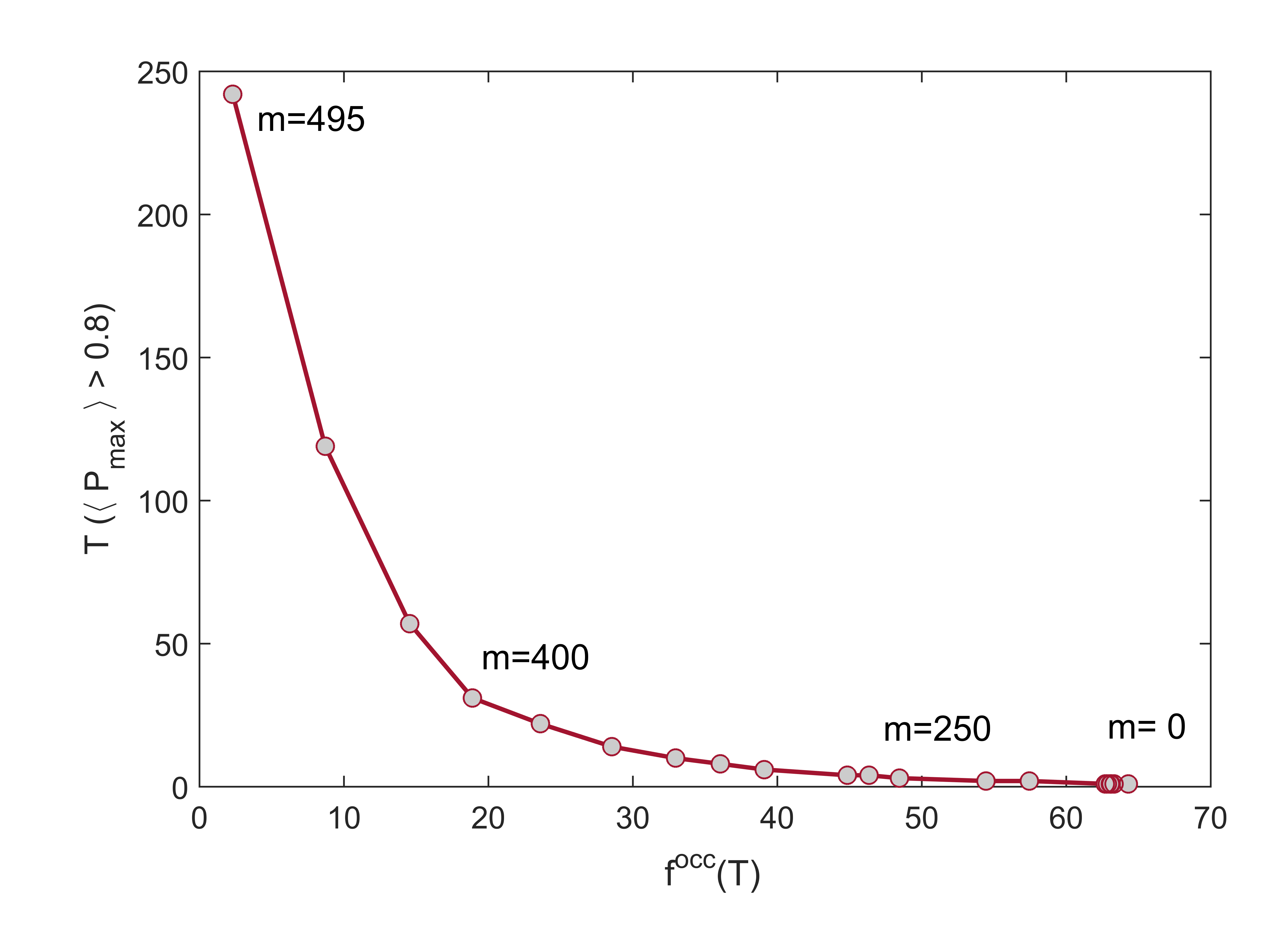}
\caption{ Trade-off in efficiency and time of convergence with Polya updating scheme. Simulation results show a clear monotonic decay for $N$ = 500.}
\label{fig:trade-off}
\end{figure}
We study this trade-off in Fig. \ref{fig:trade-off} which plots the number of time slices required by the average over maximum probabilities to reach at least 0.8 versus
the percentage of restaurants occupied. The variable in the $y$-axis represents the cost in terms of waiting time. The variable in the $x$-axis represents
the cost in terms of inefficiency of the solution (smaller percentage occupancy would be more efficient). We plot the trade-off by simulating a system of $N$ = 500
agents with the {\it Polya} updating scheme ($m$ = 50, 75, 100, $\ldots$, 475, 495). the values on the $x$-axis shows the occupancy at the time slice when $<P_{max}>$ reaches 0.8. 

The trade-off is clearly seen in terms of cost minimization. A lower waiting cost leads to higher occupancy and hence to inefficiency and vice versa.
This is a very useful feature of the model to understand the trade-off between the waiting cost to arrive at an allocation and the accuracy of the allocation.

\section{Summary}

In this paper, we study a model of distributed coordination in the context of a multi-agent, multi-choice system. We consider a game with
multiple Nash equilibrium all of which are equally likely. The basic problem is to find which equilibrium will materialize if the agent engage in repeated
interaction and how quickly can they converge to the equilibrium. Essentially, we solve the problem of equilibrium selection through
distributed coordination algorithms.

We propose a number of strategies based on different types of {\it na\"{i}ve} learning. In particular, reinforcement learning via {\it Polya's urn model}
provides a very useful benchmark. We show that the system self-organizes with very slow dynamics and transient clusters. Finally, we characterize
a trade-off between waiting cost to attain an allocation and the accuracy of the allocation. With lower waiting costs (stability is attained sooner),
efficiency of the solution is low and the opposite is also true.

This problem sheds light on complexity of equilibrium selection and may provide an useful model for multi-agent coordination and collective dynamics in general. 

\bibliographystyle{IEEEtr}

\end{document}